\documentclass{article}
\usepackage{spconf,amsmath,graphicx,multirow,comment,url,bbding}

\title{T-VECTORS: WEAKLY SUPERVISED SPEAKER IDENTIFICATION USING HIERARCHICAL TRANSFORMER MODEL}
\name{Yanpei Shi, Mingjie Chen, Qiang Huang, Thomas Hain}
\address{ Speech and Hearing Research Group\\
	Department of Computer Science, University of Sheffield\\
	\texttt{\{YShi30, mchen33, qiang.huang, t.hain\}@sheffield.ac.uk}}
\begin{document}
\ninept
\maketitle
\begin{abstract}
Identifying multiple speakers without knowing where a speaker's voice is in a recording is a challenging task. This paper proposes a hierarchical network with transformer encoders and memory mechanism to address this problem. The proposed model contains a frame-level encoder and segment-level encoder, both of them make use of the transformer encoder block. The multi-head attention mechanism in the transformer structure could better capture different speaker properties when the input utterance contains multiple speakers. The memory mechanism used in the frame-level encoders can build a recurrent connection that better capture long-term speaker features. The experiments are conducted on artificial datasets based on the Switchboard Cellular part1 (SWBC) and Voxceleb1 datasets. In different data construction scenarios (Concat and Overlap), the proposed model shows better performance comparaing with four strong baselines, reaching 13.3\% and 10.5\% relative improvement compared with H-vectors and S-vectors. The use of memory mechanism could reach 10.6\% and 7.7\% relative improvement compared with not using memory mechanism.

\end{abstract}
\begin{keywords}
Weakly Supervised Learning, Speaker Identification, Transformer.
\end{keywords}
\section{Introduction}

Speaker identification aims to identify the speakers' identities from the input audio recording \cite{variani2014deep,wang2018attention}. Conventional supervised speaker identification requires the manual labelling of the input data, where the segments and the corresponding speaker labels are manually annotated \cite{karu2018weakly}. However, it might be difficult to process a large dataset with a large number of speakers using hand annotation \cite{karu2018weakly,jia2019leveraging}.   
Instead, weakly supervised training only relies on the set of speaker labels that occur in the corresponding utterance \cite{zhou2018brief}. This kind of weakly labelled large data collections are available online \cite{karu2018weakly}. Making use of such datasets might be beneficial for training with a large amount of data \cite{shi2020weakly}.

In speech technology, weakly supervised training has been widely used. In \cite{karu2018weakly}, Karu et al.  proposed a DNN based weakly supervised speaker identification training technique, where a DNN is trained to predict the set of speaker labels using the utterance-level speaker labels. In \cite{xu2017unsupervised} and \cite{xu2018large}, Xu et al. proposed a DNN based approach and a gated convolutional neural network for multi-label audio classification.

Our previous work \cite{shi2020weakly} proposed a hierarchical attention network (H-vectors) to solve the weakly supervised speaker identification problem. However, the H-vectors proposed in \cite{shi2020weakly} might have two disadvantages: the use of the GRU in each frame-level encoder might influence the training speed and cause gradient vanishing problem, and there is no connections between each frame-level encoders, which might influence the performance.

In this work, a hierarchical network based on transformer's encoder architecture \cite{vaswani2017attention} is proposed to tackle the weakly supervised speaker identification problem. Based on our previous work \cite{shi2020weakly}, the proposed model (T-vector) contains a frame-level encoder and segment-level encoder. However, the T-vector model makes use of transformer encoder blocks in both frame-level encoder and segment-level encoder to better capture speaker information and process the sequence in parallel instead using GRU layer \cite{chung2014empirical}. The multi-head attention mechanism \cite{vaswani2017attention} used in the transformer block might better capture different speaker properties from the input multi-speaker signal.  
Additional memory mechanism between each frame-level encoder is used. The memory mechanism builds up a recurrent connection, which could reuse the hidden states from previous segments to better capture long-term information \cite{dai2019transformer}.

The rest of the paper is organized as follow: Section \ref{Model Architecture} 
presents the architecture of our approach. 
Section \ref{Experiments} depicts the data and the data construction process, the experimental setup, the baselines to be compared and implementation details.
The results are obtained and shown in Section \ref{Results}, and a conclusion is in Section \ref{Conclusion and Future Work}.

\section{Model Architecture}\label{Model Architecture}
\begin{figure}[h]
	\centering
	\includegraphics[height=14cm,width=8cm]{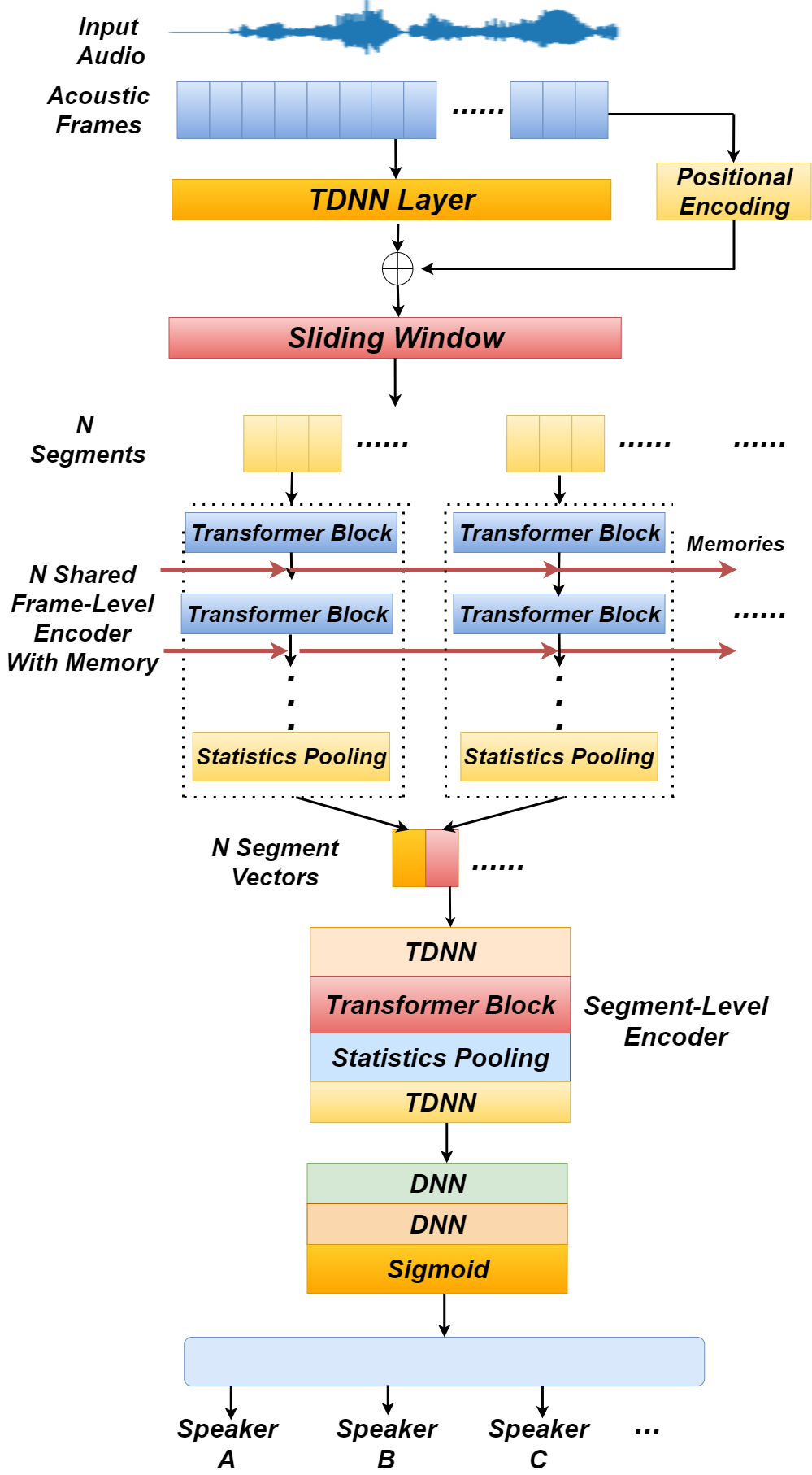}
	\caption{The architecture of the proposed model.}
	\label{proposed architecture}
\end{figure}

Figure \ref{proposed architecture} shows the architecture of the proposed model, which consists of several parts: a global TDNN layer, a positional encoding operation (sinusoidal positional encoding \cite{vaswani2017attention}), frame-level encoders,
a segment-level encoder, and two fully connected layers as a classifier.

Given the input acoustic frame vector sequence, a TDNN layer is used, and the output is then added with the positional encoding of the original input, the output is denoted as $\boldsymbol S \in \mathcal {R}^{T \times D}$ ,where $T$ represents the sequence length and $D$ is the dimension of the frequency axis.

In the frame-level encoders, 
the input sequence $\boldsymbol S$ is divided into $N$ segments: $\textbf{S} = \{\textbf{X}_1, \textbf{X}_2, \cdots, \textbf{X}_N\}$
using a sliding window with length $M$ and step $H$. Then, $N$ frame-level encoders are used, compressing $N$ segments into $N$ segment vectors $\boldsymbol V_{S_{i}}, i \in \{1,2...N\}$. Each frame-level encoder contains $L$ layers of transformer encoder block and a statistics pooling operation. Each frame-level encoder block shares weights, and connected using memories (hidden states).

After obtaining the segment vector sequence $\boldsymbol V_{S} \in \mathcal {R}^{N \times D}$, the segment-level encoder uses a TDNN layer followed by a transformer encoder block (memory is not used in segment level encoder), another TDNN layer and a statistics pooling are used to compress the segment vector sequence into a single vector that represents the whole input sequence (utterance vector). The final speaker identity classifier is constructed using a two-layer MLP followed by a sigmoid activation function \cite{ito1991representation}. 

The final speaker identities are the output vector which contains the scores (between 1 and 0) for each speaker. The model is trained using binary cross entropy loss (shown in Eq \ref{bce}) \cite{xu2017unsupervised,shi2020weakly}: 
\begin{equation}\label{bce}
E_{bce} = - \sum_{n=1}^{BN}||\boldsymbol {Y_{n}} \log \boldsymbol {\hat{Y_{n}}} + (1- \boldsymbol {Y_{n}}) \log (1-\boldsymbol {\hat{Y_{n}}})||
\end{equation}
, where $\boldsymbol {\hat{Y_{n}}}$ denotes the predicted score vector, $\boldsymbol Y_{n}$ denotes the reference label vector, $BN$ denotes the batch size.

\subsection{Transformer Encoder Block}

\begin{figure}[h]
	\centering
	\includegraphics[height=6.5cm,width=8cm]{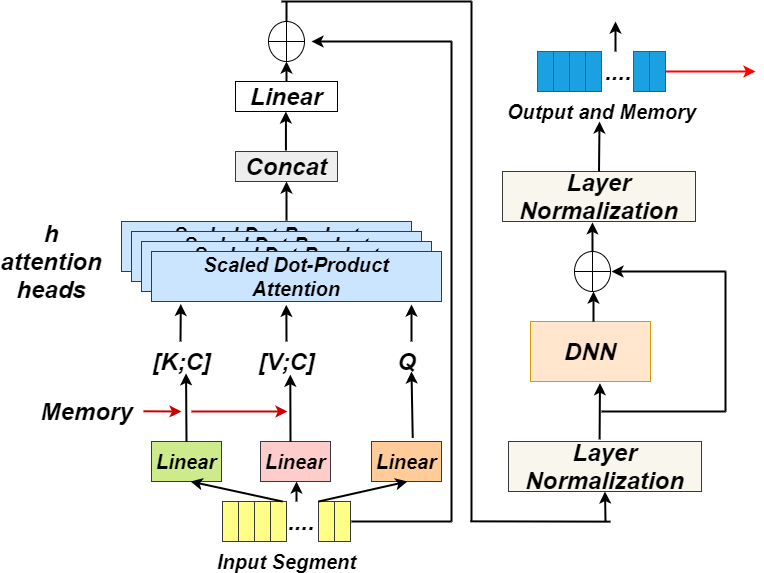}
	\caption{The architecture of the transformer encoder block with memories \cite{vaswani2017attention,dai2019transformer}.}
	\label{encoder_block}
\end{figure}

Figure \ref{encoder_block} shows the architecture of each transformer encoder block \cite{vaswani2017attention}. For each block. A multi-head attention layer is used for the input segment \cite{vaswani2017attention}. Layer normalization \cite{ba2016layer} is used after the residual connection. The output is then fed into a DNN layer, and the same layer normalization is used after the skip connection. The output is then used as the input to the next block and the memory for the next frame-level encoder. 
\subsection{Multi-Head Attention With Memories}
\begin{equation}\label{multihead}
\begin{aligned}
&MultiHead([Q;C],[K;C],V) = Concat(head_{1,2,...h})W^{O}\\
&head_{i} = Attention(QW^{Q}_{i},[K;C]W^{k}_{i},[V;C] W^{V}_{i})\\
&Attention = softmax(\frac{Q[K;C]^{T}}{\sqrt{D}})[V;C]
\end{aligned}
\end{equation}

Eq \ref{multihead} shows the computation process of the multi-head attention with memories (shown in Figure \ref{encoder_block}) \cite{dai2019transformer}. The input segment is transformed into Queries (Q), Keys (K) and Values(V) using linear transformation ($K,Q,V \in \mathcal {R}^{M \times D}$) \cite{vaswani2017attention}. Then, memory (C, has the same dimensionality as K, Q and V) from the last block is concatenated with K and V respectively, results in $[K;C], [Q;C] \in \mathcal {R}^{2M \times D}$. For each attention head, $W^{Q}_{i}, W^{Q}_{i}, W^{Q}_{i} \in \mathcal {R}^{D_{k} \times M}$ are the parameter matrices, $D_{k} = \frac{D}{h}$, $h$ is the number of attention heads. The results for each attention heads are concatenated together and $W^{O} \in \mathcal {R}^{D \times M}$ is used to fuse the output of each attention head into a single output. The values of the initial memories are set to zeros when processing the first segment of the input sequence, the gradients are not computed for the all of the memories during training process \cite{vaswani2017attention,dai2019transformer}. 

The reasons of making use of the multi-head attention and the memory mechanism are manifold. In weakly supervised speaker identification task, the input utterance might contain multiple speakers with overlaps. The use of multiple attention heads might make the model focus on different speaker properties thus better predict the speaker identities. One speaker property might occur in every part of the input sequence, thus the use of memories (hidden states) might better capture long-term speaker features.

\begin{figure*}[h]
	\centering
	\includegraphics[height=7.5cm,width=18cm]{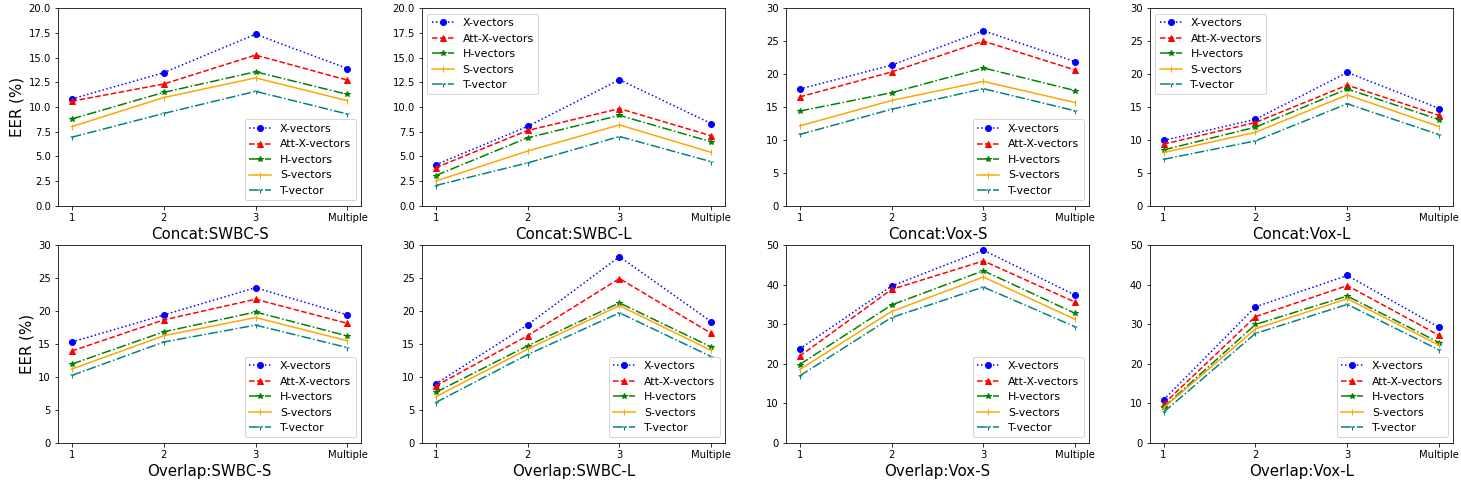}
	\caption{The results obtained using the five models (X-vectors, Attentive X-vectors, H-vectors, S-vectors, T-vectors) in different test conditions (1, 2, 3 or multiple speakers) on the eight designed datasets (SWBC-S, SWBC-L, Vox-S and Vox-L) and scenarios (Concat and Overlap). For all of the figures, the x-axis represents the number of speakers in test utterance. In T-vector model, the window size $M$ is 20 frames, the step size $H$ is 10 frames.}
	\label{all_results}
\end{figure*}

\section{Experiments}\label{Experiments}
\subsection{Data}

For this work, as there is no data for weakly supervised speaker identification task, and to compare with our previous published results, the data construction process is the same as that in our previous work \cite{shi2020weakly}. 

More specifically, eight different artificial augmented datasets are generated using Switchboard Cellular Part 1 (SWBC) \cite{swb} and Voxceleb1 \cite{nagrani2017voxceleb} corpora. There are two scenarios in the eight generated datasets: Concat scenario, where multiple speakers in the input utterance are concatenated without an overlap; Overlap scenario, where the speakers' voices are completely overlapped. For each of the two data construction scenarios, four datasets are generated: ``SWBC-S'' and ``SWBC-L'' (based on SWBC dataset), where ``S'' and ``L'' represents small and large version; ``Vox-S'' and ``Vox-L'' (based on Voxceleb1 dataset), which represents small and large version of the generated Voxceleb1 based dataset. The small and large version of the generated datasets are used to evaluate the robustness of the proposed model. More details of the eight generated datasets can be found in \cite{shi2020weakly}. The number of speakers in the input utterances are randomly chosen from one to three in all of the eight dataset, 20-dimensional MFCCs \cite{tiwari2010mfcc} are used as the input acoustic features, the utterance length is set to 5 seconds.

\subsection{Experiment Setup}

Four strong baselines are chosen to compare with the proposed model: X-vectors \cite{snyder2018x}, Attentive X-vector (Att-Xvector) \cite{zhu2018self,okabe2018attentive,wang2018attention, rahman2018attention}, H-vector \cite{shi2020weakly,shi2020h} and S-vector \cite{katta2020s}.
X-vectors is TDNN based model, which contains a TDNN based frame-level feature extractor, a statistics pooling operation and a segment-level feature extractor. Att-X-vector contains an additional global self attention mechanism to allocate different weights for different frames before the statistics pooling operation. 
H-vector is based our previous work \cite{shi2020weakly,shi2020h}, where the hierarchical structure is similar to the proposed model, but the frame-level encoder and segment-level encoder are based on TDNN and GRU structure, and there is no connections between different frame-level encoders. 
S-vector \cite{katta2020s} is based on transformer's encoder, where the frame-level feature extractor in X-vectors is replaced with a transformer's encoder. The input sequence is processed globally, hierarchical structure is not used in this baseline. 
The proposed approach is denoted as ``T-vector", similar to our previous work \cite{shi2020weakly}, the window length $M$ is set to 20 frames, and the step length $H$ is set to 10 frames.

In this work, equal error rate (EER) \cite{cheng2004method,murphy2012machine} is used as the evaluation metric, as it is widely used in multi-label audio tagging \cite{xu2017unsupervised}. The EER is defined as the point when the false negative (FN) equals to the false positive rate (FP) rate. EER is computed for each individual input and averaged across the whole test set \cite{cheng2004method}.

\subsection{Implementation}
\vspace*{-4mm}

\begin{table}[h]
	\renewcommand\arraystretch{1.2}
	\setlength{\tabcolsep}{1.8mm}
	\centering  
	\caption{The architecture of the proposed network architecture, where K denotes the total number of speakers, T represents the whole sequence length, Transformer represents one transformer encode block.}\label{model_summary}
	\begin{tabular}{c|c|c|c}
		\hline
		Level& Model & Input & Output  \\
		\hline
		Global & TDNN & (T,20) & (T,D)\\
		\hline
		\multirow{2}{*}{Frame-Level}&
		4xTransformer & (M,D) & (M,D)\\
		&Statistics Pooling & (M,D)& (1,2D)\\
		\hline
		
		\multirow{4}{*}{Segment-Level}&
		TDNN & (N,2D) & (N,D)\\
		&Transformer & (N,D)& (N,D)\\
		&TDNN& (N,D) & (N,1500)\\
		&Statistics Pooling & (N,1500)& (1,3000)\\
		\hline

		\multirow{2}{*}{Utterance-Level}&
		DNN (512) & (1,3000) & (1,512)\\
		&DNN (K)& (1,512)& (1,K)\\
		\hline
\end{tabular}
\label{model_sum}
\end{table}

\begin{table*}[h]
	\renewcommand{\multirowsetup}{\centering}  
	\renewcommand\arraystretch{1.2}
	\setlength{\tabcolsep}{3mm}
	\centering  
	\caption{The obtained results of using the memory mechanisms or not for the T-vector model, the window size $M$ is changing from 20 to 30 frames.}
	\begin{tabular}{c|c|c|c|c|c|c}
		\hline
		\multirow{2}{*}{\textbf{Data Type}}& \multirow{2}{*}{\textbf{Memory}} & \multirow{2}{*}{\textbf{Window Size}} &\multicolumn{3}{c}{\textbf{EER (\%)}}\\
		
		\cline{4-7}
		& & & SWBC-S& SWBC-L&Vox-S&Vox-L\\
		\hline
		\multirow{6}{*}{\textbf{Concat}}&
		\multirow{3}{*}{\textbf{With}}&
		20&9.28&4.46&14.41&10.79\\
		& &25&\textbf{8.97}&\textbf{4.04}&\textbf{13.97}&\textbf{10.25}\\
		& &30&9.05&4.20&14.30&10.49\\
		\cline{2-7}
		&\multirow{3}{*}{\textbf{Without}}&
		20&10.31&5.52&16.02&11.64\\
		& &25&10.04&5.18&15.54&10.98\\
		& &30&10.34&5.37&15.85&11.47\\
		\hline

		\multirow{6}{*}{\textbf{Overlap}}&
		\multirow{3}{*}{\textbf{With}}&
		20&14.48&13.10&29.38&23.51\\
		& &25&\textbf{14.07}&12.96&\textbf{28.91}&23.25\\
		& &30&14.13&\textbf{12.77}&29.04&\textbf{23.08}\\
		\cline{2-7}
		&\multirow{3}{*}{\textbf{Without}}&
		20&15.85&14.25&31.04&24.48\\
		& &25&15.25&13.99&30.24&23.80\\
		& &30&15.71&13.75&30.09&23.66\\
		\hline

	\end{tabular}
	\label{memory_res}
\end{table*}

Table \ref{model_sum} shows the parameter configuration of the proposed network architecture. The value of $D$ is set to 512, and the dimension of the DNN within the transformer encoder block is set to 2048, the number of the attention heads $h$ is set to 4 in all transformer layers. 
The TDNN layers in both frame-level and segment-level encoder operate at the current time step.
The adam optimiser \cite{Kingma2014AdamAM} is used for all experiments with $\beta_1=0.95$, $\beta_2=0.999$, and $\epsilon= 10^{-8}$. The initial learning rate is $10^{-4}$.

\section{results}\label{Results}

Figure \ref{all_results} shows the results obtained using the five models (X-vectors, Attentive X-vectors, H-vectors, S-vectors, T-vectors) in different test conditions. The T-vector model performed better than the four baselines in all of the test conditions. The T-vector model shows robustness when the training data is small, reaching 13.3 \% and 10.56 \% relative improvement than H-vectors and S-vectors in SWBC-S dataset in Concat scenario. Compared with the results obtained with H-vectors (using the similar hierarchical structure), the improvement of T-vectors comes from the use of multi-head attention and the memory mechanism. Compared with the TDNN and GRU layers used for H-vectors, the multi-head attention mechanism used in T-vectors might better capture multiple speaker properties when the input utterance contains multiple speakers, as the different attention heads might focus on different speaker features. The Memory mechanism used for T-vectors connects different frame-level encoders, which might better capture long-term speaker features. When the length of the input utterance (5 seconds in this work) is large, the memory mechanism can make the model better capture long-term features.

Compared with the results obtained by S-vectors, the improvement of T-vectors comes from the use of hierarchical structure with the memory mechanism. Instead of processing the whole input sequence in S-vectors, the T-vector makes use of the hierarchical structure that can better capture local and global features. The memory mechanism grantees the information of each speaker could be shared in each frame-level encoders.   
Compared with the X-vectors and Attentive X-vector baselines, the improvement of T-vectors comes from the use of multi-head attention mechanism, which could better capture overlapped speaker information than TDNN layers.

Among all of the test conditions, the best results are obtained when the number of speakers in each utterance is one, and the worse case is when each utterance contains three speakers. This might due to the difficulty of the test conditions that it is difficult to distinguish different speakers overlapped in the input utterance. A similar reason also occurs in the two different data construction scenarios (Concat and Overlap).

Table \ref{memory_res} shows the obtained results of T-vectors when using the memory mechanism or not with different window size ($M$). Among all of the results, the best results in both Concat and Overlap scenarios are obtained when using the memory mechanism. The use of the memory mechanism could significantly improve the performance of T-vectors, reaching 10.6\% and 7.7\% relative improvement in the two scenarios in ``SWBC-S'' dataset when the window size is 25 frames. It shows that reusing the information from the previous segments could make the model better capture the long-term speaker information. In addition, in most of the window size settings, the best results are obtained when the window size is 25 frames. This shows a reasonable window size could influence the performance. When the window size becomes larger (30 frames), a single frame-level encoder might not be able to better capture the speaker information within the current segment.

\section{Conclusion}\label{Conclusion and Future Work}
In conclusion, a hierarchical network with transformer encoder block and memory mechanism is proposed to solve the weakly supervised speaker identification problem. The proposed model makes use of the transformer encoder block in both frame-level and segment-level encoder. Memory mechanism is used to connect different frame-level encoders to better capture long-term memories. The experiments are done with different test conditions and different amount of training data. The obtained results show that the proposed T-vector model out-performs the four strong baselines.

In the future work, different positional encoding techniques will be investigated. The number of speakers in one input utterance will be increased and larger dataset such as Voxceleb2 will be used.

\begin{center}
	\large{\textbf{Acknowledgement}}
\end{center}
This work was in part supported by Innovate UK Grant number 104264.

\newpage
\small
\bibliographystyle{IEEEbib}
\bibliography{refs}

\end{document}